# Modeling and FEM-based Simulations of Composite Membrane based Circular Capacitive Pressure Sensor

Rishabh Bhooshan Mishra, S Santosh Kumar, Ravindra Mukhiya

Smart Sensor Area, CSIR-Central Electronics Engineering Research Institute (CEERI), Pilani, Rajasthan, India – 333031
santoshkumar.ceeri@gmail.com

**Abstract.** In Micro-electro-mechanical Systems (MEMS) based pressure sensors and acoustic devices, deflection of a membrane is utilized for pressure or sound measurements. Due to advantages of capacitive pressure sensor over piezoresistive pressure sensors (low power consumption, less sensitive to temperature drift, higher dynamic range, high sensitivity), capacitive pressure sensors are the 2nd largest useable MEMS-based sensor after piezoresistive pressure sensors. We present a normal capacitive pressure sensor, for continuous sensing of normal and abnormal Intraocular Pressure (IOP). The composite membrane of the sensor is made of three materials, i.e., Si, $SiO_2$ and $Si_3N_4$. The membrane deflection, capacitance variation, mechanical sensitivity, capacitive sensitivity and non-linearity are discussed in this work. Mathematical modeling is performed for analytical simulation, which is also compared with Finite Element Method (FEM) simulations. MATLAB$^®$ is used for analytical simulations and CoventorWare$^®$ is used for FEM simulations. The variation in analytical result of deflection in membrane w.r.t. FEM result is about 7.19%, and for capacitance, the variation is about 2.7% at maximum pressure of 8 kPa. The non-linearity is about 4.2492% for the proposed sensor for fabrication using surface micro-machining process.

## 1  Introduction

Micro-machined sensors are integrated with electrical interface to make electromechanical systems. MEMS sensors are systems which interact with measurand (like displacement, acceleration, flow, pressure and temperature etc.) and then converts it to an electrical signal, which is used to analyze the measurand so that further controlling, minoring and/or alarming actions can take place. ICs are used to perform signal-conditioning so that obtained signal can be used for further decision making/transmission/communication [1]. The market of MEMS based devices is growing at a fast rate from several years. In sensors domain, MEMS based inertial sensors, pressure sensors and micro-actuators have large application not only for consumer electronics, defense systems and automobile application but also for bio-medical ap-plication. In some biomedical applications, the devices can be implanted in living-being, and then bio-medical signal can be transmitted using wireless communication technology i.e. telemetry [2-4].

MEMS based pressure sensors have replaced conventionally available pressure measuring devices like bourdon tubes, bellows, diaphragms, capsules and various vacuum measuring devices (such as Pirani gauge, McLeod gauge etc.). In conventional pressure measuring devices, pressure causes mechanical movement that rotates the pointers/dial. However, MEMS devices directly convert input pressure into corresponding electrical signal. MEMS pressure sensors are fabricated on silicon wafers (SOI, double SOI or single crystal silicon wafers) using surface, bulk or a combination of these two micro-machining techniques. Utilizing the above-mentioned technologies, the diaphragms are released which acts as a sensing element [5].

Increase or fluctuation in IOP can be the cause of glaucoma. According to World Health Organization (WHO), glaucoma is one of the major cause of blindness. Glaucoma causes irreversible eye disease, which damages the optical nerves. The normal range of IOP is 1.6 kPa – 2.8 kPa. Therefore, the accurate measurement in early stage can save the eye from permanent blindness [2-3].

The normal mode capacitive pressure sensors have a fixed plate and a movable (usually conductive) plate/membrane, which are separated by a medium, i.e., vacuum, air or dielectric materials. The capacitance variation can be obtained by following three techniques (usually first technique is utilized for designing MEMS capacitive pressure sensor):

- Changing the separation gap between parallel plates.
- Changing the overlapping area between parallel plates.
- Movement changing in dielectric materials which is filled between plates.

This paper presents mathematical/theoretical modeling and FEM simulation of composite membrane based normal mode capacitive pressure sensor for IOP measurement (0 – 8 kPa). The obtained results are also compared with one of our previous works, in which the modeling and FEM simulation of normal mode capacitive pressure sensor is carried out for the same application. In that work, only the silicon material of Young modulus of elasticity 169 GPa and poisson ratio of 0.066 is used as diaphragm material. For different diaphragm thicknesses after optimization, comparison of theoretical and simulation results is presented [3]. In this presented work, the composite membrane is made of silicon (with same material properties), silicon dioxide (Young's modulus of elasticity 70 GPa and poisson ratio 0.17) and silicon nitride (Young's modulus of elasticity 222 GPa and poisson ratio 0.27).

## 2      Mathematical Modeling of Sensor

### 2.1    Membrane deflection and Capacitance variation

After application of pressure P on thin, clamped and flat circular diaphragm of radius $a$, thickness $t$ which made of homogeneous, isotropic and elastic material with

Young's modulus of rigidity E and Poisson ratio ν, the diaphragm deflection can be given by [3]:

$$w(r) = \frac{Pa^4}{64\,D}\left[1-\left(\frac{r}{a}\right)^2\right]^2 \qquad (1)$$

Here, D (flexural rigidity) can be given by:

$$D = \frac{Et^3}{12(1-v^2)} \qquad (2)$$

The base capacitance of sensor can be given by:

$$C_{base} = \frac{\varepsilon A}{d} \qquad (3)$$

Here, A, d and ε are overlapping area between plates, separation gap and permittivity of medium, respectively.

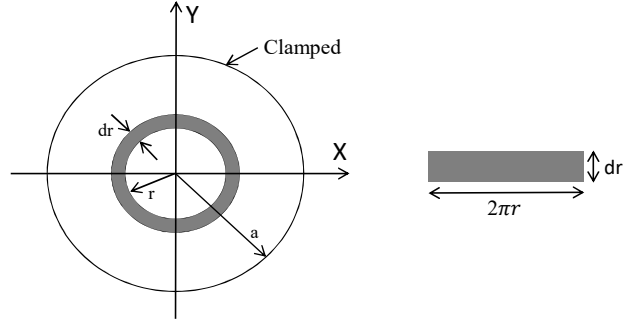

**Fig. 1.** Circular clamped diaphragm depicting an annulus is taken from composite membrane.

To find the capacitance of parallel plate capacitor after pressure application, cut an annulus at radius r of width dr from deflected clamped circular diaphragm, as shown in Fig. 1. The capacitance due to this annulus (a small element) can be given by:

$$\partial C_d = \frac{\varepsilon(2\pi r\ )}{d-w(r)} \qquad (4)$$

After performing the integration over whole area of sensor, the capacitance can be given by:

$$C_d = \int_0^a \frac{\varepsilon(2\pi r dr)}{d-\frac{Pa^4}{64\,D}\left[1-\left(\frac{r}{a}\right)^2\right]^2} \qquad (5)$$

After solving the above equation [3]:

$$C_w = 4\pi\varepsilon\sqrt{\frac{D}{Pd}}\,\ln\left|\frac{a^2\sqrt{P}+8\sqrt{dD}}{a^2\sqrt{P}-8\sqrt{dD}}\right| \qquad (6)$$

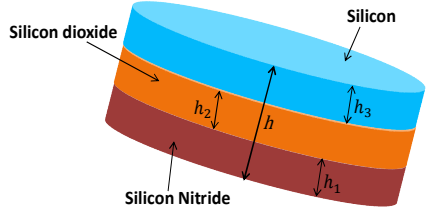

**Fig. 2.** Composite membrane made of three different materials.

If the diaphragm material is made of three different materials silicon, silicon dioxide and silicon nitride [$Si_3N_4(E_1, v_1, h_1)/SiO_2(E_2, v_2, h_2)/Si(E_3, v_3, h_3)$], as shown in Fig. 2, then flexural rigidity of composite membrane can be given by [5]:

$$D_c = \frac{E_1[(h-e)^3-(h-e-h_1)^3]}{3(1-v_1^2)} + \frac{E_2[(h-e-h_1)^3-(h_3-e)^3]}{3(1-v_2^2)} + \frac{E_3[(h_3-e)^3+e^3]}{3(1-v_3^2)} \quad (7)$$

here, e is the neutral plane (the plane within the composite plate/membrane which is not in tension, compression or stress when there is application of pressure on the plate/membrane), this can be given by [5]:

$$e = \frac{\frac{E_1}{1-v_1}h_1(h_1+2h_2+2h_3)+\frac{E_2}{1-v_2}h_2(h_2+2h_3)+\frac{E_2}{1-v_3}h_3^2}{2\left[\frac{E_1}{1-v_1}h_1+\frac{E_2}{1-v_2}h_2+\frac{E_2}{1-v_3}h_3\right]} \quad (8)$$

Then deflection in the composite membrane, of radius L and flexural rigidity and thickness can be given by:

$$w_{composite}(r) = \frac{Pa^4}{64\,D_c}\left[1-\left(\frac{r}{a}\right)^2\right]^2 \quad (9)$$

The capacitance variation due to pressure application is obtained by modifying the Eq. (4) and can be given by:

$$C_{w\_composite} = 4\pi\varepsilon\sqrt{\frac{D_c}{Pd}} \ln\left|\frac{a^2\sqrt{P}+8\sqrt{dD_c}}{a^2\sqrt{P}-8\sqrt{dD_c}}\right| \quad (10)$$

### 2.2 Sensitivity and Non-linearity

The mechanical sensitivity is a useful parameter, if maximum membrane deflection varies in different designs. The mechanical sensitivity is slope of maximum deflection versus pressure range curve. The mechanical sensitivity of composite membrane can be given by [2-3]:

$$S_{mech,composite} = \frac{a^4}{64\,D_c} \quad (11)$$

The capacitive sensitivity of composite membrane based sensor is obtained by the ratio of change in capacitance and applied pressure range [2-4]:

$$S_{cap,composite} = \frac{C_{max}-C_{min}}{P_{max}-P_{min}} \quad (12)$$

Mechanical and Capacitive sensitivity both are terms which define the performance and specifications of sensors.

The non-linearity of the sensor, at particular point, can be defined by [6]:

$$NL_i(\%) = \frac{C_{d\_composite}-C_{d\_composite}\times\frac{P_i}{P_m}}{C_{d\_composite}} \times 100 \quad (13)$$

here, $P_i$ is applied pressure at any point on the calibrated curve, $P_m$ is maximum pressure, $C_{d\_composite}(P_i)$ is capacitance at particular point on calibrated curve and $C_{d\_composite}(P_m)$ is capacitance at maximum pressure.

## 3  Results and Discussion

The sensor must have good sensitivity, minimum non-linearity, low power consumption, robustness and small size. Before fabrication of the pressure sensor several steps needs to be carried out like designing, mathematical formulation, verification of mathematical modeling, analytical simulations and comparison of analytical simulation with FEM simulation. According to the mathematical formulation and analytical simulation, the optimized design of sensor can be obtained. However, in modeling and analytical simulation, several assumptions need to be taken. Therefore, the FEM simulation is performed to analyze the behavior of the sensor in a practical situation. The optimization of device using fabrication runs is complex, costly, time taking and needs more effort. Hence, it is highly desirable to have accurate and precise design and modeling. Considering the practical scenario, in this work, we have modeled composite membrane and verified the modeling with FEM simulations.

In a capacitive pressure sensor, base capacitance must be in pF range and change must be in fF range for the capacitance measurement to be performed, effectively and efficiently. Therefore, we should try to increase base capacitance and change in capacitance by choosing appropriate and optimized design parameters for sensors. If over-lapping area is increased for a particular thickness of membrane then deflection of membrane increases, so that mechanical sensitivity as well as capacitive sensitivity increases. If thickness of membrane is increased by keeping over-lapping area between plates constant/same then deflection in diaphragm decreases, so that both mechanical and capacitive sensitivity decreases.

### 3.1  Analytical design for membrane deflection

The comparison of diaphragm deflection in 6 μm thick silicon diaphragm and composite diaphragm of same thickness is performed. In all the designs, composite membranes have a thickness of 6 μm (t). Silicon thickness is kept 5.3 μm thick and thickness of $SiO_2$ and $Si_3N_4$ are varied as shown in the Table 1. The deflection in various composite diaphragms and in Silicon diaphragm with same overall thickness is shown in Fig 3. The radius of the diaphragm is 360 μm and applied pressure is 8 kPa. The flexural rigidity and maximum deflection in membrane of these four membranes are indicated in Table 1.

**Table 1.** Flexural rigidity and maximum deflection for different models.

| Model No. | Composite Diaphragm Thicknesses Specification (μm) | Flexural Rigidity (Pa.cm$^3$) | Maximum Deflection (μm) |
|---|---|---|---|
| 1. | $t = 6$ | 3.0553 | 0.68717 |
| 2. | $h_1 = 0.2, h_2 = 0.5, h_3 = 5.3$ | 2.7556 | 0.76192 |
| 3. | $h_1 = 0.15, h_2 = 0.55, h_3 = 5.3$ | 2.6948 | 0.77910 |
| 4. | $h_1 = 0.1, h_2 = 0.6, h_3 = 5.3$ | 2.6312 | 0.79793 |

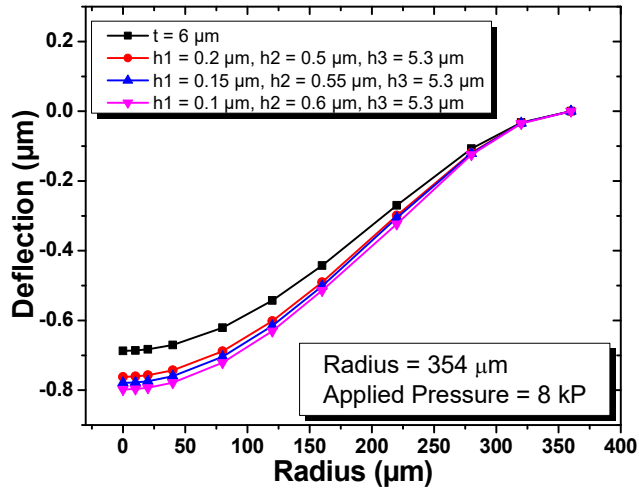

**Fig. 3.** Membrane deflection of different thicknesses.

According to Table 1 and Fig. 3, if flexural rigidity of membrane is larger, deflection is lesser; therefore, the mechanical sensitivity is less. Since the flexural rigidity of Model No.4 is less and deflection is maximum, hence, Modal No.4 is chosen for FEM simulation. The radius is modified according to separation, so that sensor can be fabricated for IOP measurements using appropriate process flow.

### 3.2    FEM simulations and comparison with analytical results

While performing FEM simulations using CoventorWare®, composite membranes are merged. Then membrane edges are clamped and bottom plate is fixed. The pressure is applied on the top of membrane. The 'Tetrahedron' meshing type of Parabolic element order is used to mesh the model of sensor with element size = 5.

The MemMech module and 'Mechanical' type of physics is utilized for mechanical analysis, i.e., for obtaining deflection in membrane.

The FEM result of composite membrane deflection of thicknesses $h_1 = 0.1\ \mu m, h_2 = 0.6\ \mu m, h_3 = 5.3\ \mu m$ and radius 354 μm is shown in Fig. 4.

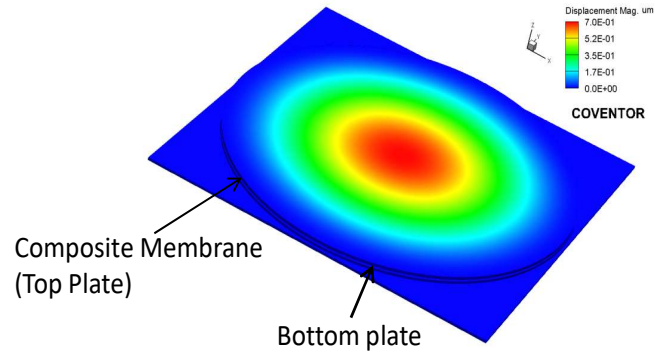

**Fig. 4.** FEM result of Deflection in Composite membrane based Circular shaped Capacitive Pressure Sensor at 8 kPa Pressure using CoventorWare®.

The comparison of analytical and FEM results are close to each other which is shown in Fig. 5. The variation in analytical and FEM results for membrane deflection increases, as applied pressure increases.

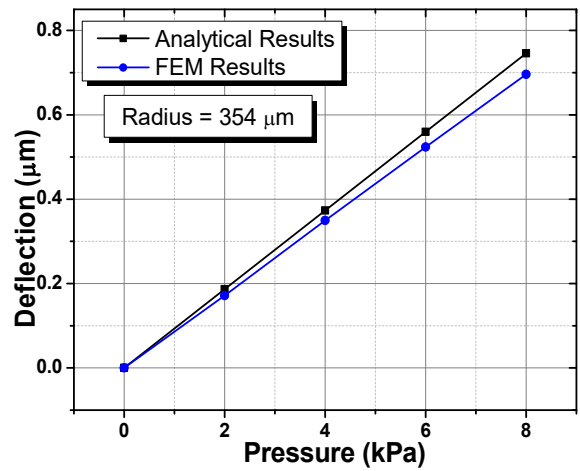

**Fig. 5.** Analytical and FEM results of maximum deflection w.r.t. pressure variation.

The MemElectro module and 'Electrostatics' physics is utilized for obtaining base capacitance. The bottom plate is grounded and 1 volt voltage is applied on the top conductor. To obtain the capacitance at a particular applied pressure, CoSolveEM module is utilized in which Surface_BCs and DC_ConductorBCs are used to define the boundary conditions. In CoSolveEM module, mechanical as well as electrical analysis can be performed at a time.

The base capacitance of sensor according to analytical simulation is 1.1619 pF and 1.206062 pF according to FEM simulation. The analytical and FEM result of capacitance variation w.r.t. applied pressure for membrane with radius of 354 μm, is shown in Fig. 6.

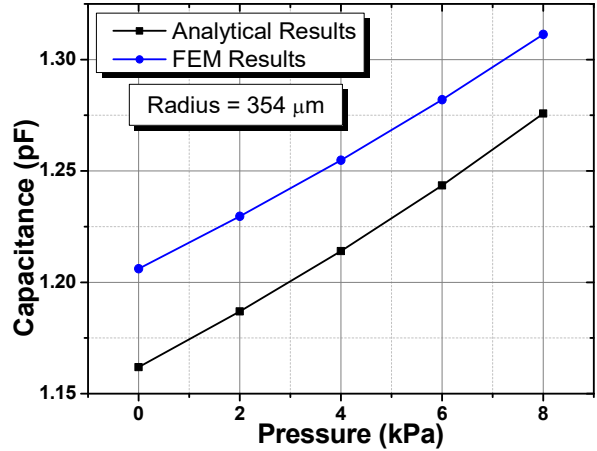

**Fig. 6.** Analytical and FEM results of Capacitance variation w.r.t. pressure variation.

The mechanical sensitivity of sensor according to analytical and FEM simulation is 93.25625 nm/kPa and 86.999075 nm/kPa, respectively. The capacitive sensitivity according to analytical and FEM simulations is 14.2375 fF/kPa and 13.1535 fF/kPa, respectively.

## 4  Conclusion

The deflection, capacitance variation and sensitivity (mechanical and capacitive both) of composite membrane based capacitive pressure sensor is discussed with mathematical modeling and simulations. The membrane of proposed design is made of three different materials namely Silicon, Silicon dioxide and Silicon Nitride with thicknesses of 5.3 μm, 0.6 μm and 0.1 μm, respectively. The deflection curve w.r.t. applied pressure is obtained linear which validates the Hook's Law. The analytical and FEM simulation result of capacitance variation w.r.t. applied pressure is also determined. The sensor has 4.2492% non-linearity. The variation in FEM results with analytical result in membrane deflation curve is due to Kirchhoff's assumptions which have been considered in mathematical modeling/formulation. In mathematical formulations and analytical simulations of base capacitance and capacitance variation, the fringing field, parasitic effects plus dielectric constant of silicon oxide and silicon nitride (i.e. 3.9 and 8, respectively) are not considered which are leading to the variation in analytical and FEM results of capacitance variation w.r.t. applied pressure.

**Acknowledgement**

The authors acknowledge the Director, CSIR-CEERI, Pilani, Rajasthan for his guidance and support. We are also thankful to valuable discussion with Dr. Ankush Jain (Scientist, Process Technologies Group) of CSIR-CEERI.